\documentclass[12pt,preprint]{aastex}

\slugcomment{draft of \today}
\shorttitle{Effects of Fluid Composition}
\shortauthors{Chattopadhyay and Ryu}

\def\etal{{\em et al.} }
\def\eg{{\it e.g.,} }
\def\egn{{\it e.g.,}}
\def\ie{{\it i.e.,} }

\begin{document}

\title{Effects of Fluid Composition on Spherical Flows around Black Holes}

\author{Indranil Chattopadhyay\altaffilmark{1} and Dongsu Ryu\altaffilmark{2*}}
\altaffiltext{1}{ARIES, Manora Peak, Nainital-263129, Uttaranchal, India:
indra@aries.ernet.in}
\altaffiltext{2}{Department of Astronomy and Space Science, Chungnam National
University, Daejeon 305-764, South Korea: ryu@canopus.cnu.ac.kr}
\altaffiltext{*}{Corresponding Author}

\begin{abstract}
Steady, spherically symmetric, adiabatic accretion and wind flows
around non-rotating black holes were studied for fully ionized,
multi-component fluids, which are described by a relativistic
equation of state (EoS).
We showed that the polytropic index depends on the temperature as well
as on the composition of fluids, so the composition is important
to the solutions of the flows.
We demonstrated that fluids with different composition can produce
dramatically different solutions, even if they have the same sonic
point, or they start with the same specific energy or the same
temperature.
Then, we pointed that the Coulomb relaxation times can be longer than
the dynamical time in the problem considered here, and discussed the
implication.
\end{abstract}

\keywords{accretion, accretion disks --- black hole physics --- hydrodynamics
--- relativity}

\section{Introduction}

It is generally inferred from observations that the matter falling
onto black holes is of very high temperature, both in microquasars
\citep{cetal03} as well as in AGNs \citep{rc00}.
The electron temperature around 10$^9$ K and/or the proton temperature
around 10$^{12}$ K or more are accepted as typical values within few tens
of the Schwarzschild radius, $r_s$, of the central black holes.
Moreover, the general theory of relativity demands that the matter
crosses the black hole horizon with the speed of light ($c$).
In other words, close to black holes, the matter is relativistic in terms
of its bulk speed and/or its temperature.
On the other hand, at large distances away from black holes, the matter
should be non-relativistic.

It is also inferred from observations that the astrophysical jets around
black hole candidates have relativistic speeds \citep{b03}.
Since the jets originate from the accreting matter very close to black
holes, their base could be very hot.
At a few hundred Schwarzschild radii above the disc plane, they can
expand to very low temperatures but very high speeds (Lorentz factor
${\gamma}\ {\ga}$ a few).
And as the fast moving matter of the jets hits the ambient medium and
drastically slows down to form shocks and hot spots, once again the
thermal energy increases to relativistic values though the bulk velocity
becomes small.
Relativistic flows are inferred for gamma-ray bursts (GRBs) too.
In the so-called collapsar model scenario \citep{w93}, the collimated
bipolar outflows emerge from deep inside collapsars and propagate
into the interstellar medium, producing GRBs and afterglows.
In such model, these collimated outflows are supposed to achieve
Lorentz factors ${\gamma}\ {\ga}\ 100$.

It is clear in the above examples that as a fluid flows onto a black hole
or away from it, there are one or more transitions from the non-relativistic
regime to the relativistic one or vice-versa.
It has been shown by quite a few authors that to describe such
trans-relativistic fluid, the equation of state (EoS) with a fixed adiabatic
index $\Gamma$ ($=c_p/c_v$, the ratio of specific heats) is inadequate and
the relativistically correct EoS \citep{c38,s57} should be used
\citep[\egn][]{t48,metal05,rcc06}.

A fluid is said to be thermally relativistic, if its thermal energy is
comparable to or greater than its rest mass energy, \ie if $kT\ \ga\ mc^2$.
The thermally non-relativistic regime is $kT \ll mc^2$.
Here, $T$ is the temperature, $k$ is the Boltzmann constant, and $m$ is
the mass of the particles that constitute the fluid.
So it is not just the temperature that determines a fluid to be thermally
relativistic, but it is the ratio, $T/m$, that determines it.
Therefore, together with the temperature, the composition of the fluid
(\ie either the fluid is composed of electron-positron pairs, or
electrons and protons, or some other combinations) will determine whether
the fluid is in the thermally relativistic regime or not.

The study of relativistic flows around compact objects including black
holes was started by \citet{m72}.
It was basically recasting the transonic accretion and wind solutions around
Newtonian objects obtained by \citet{b52} into the framework of the general
theory of relativity.
Since then, a number of authors have addressed the problem of relativistic
flows around black holes, each focusing on its various aspects
\citep[\egn][]{bm76, ft85, c96, d01, d02, metal04, betal06, fk07,
metal07}.
Barring a few exceptions \citep[\egn][]{bm76,metal04}, most of these
studies used the EoS with a fixed ${\Gamma}$, which, as we have noted,
is incapable of describing a fluid from infinity to the horizon.
\citet{bm76} for the first time calculated the spherical accretion and
wind solutions around Schwarzschild black holes, while using an
approximate EoS for the single-component relativistic fluid \citep{m71}.
\cite{metal04} modified the EoS used by \citet{bm76} to obtain
thermally driven spherical winds with relativistic terminal speeds.
However, there has been no extensive study of the effects of fluid
composition on the solutions of transonic flows around black holes.
We in this paper investigate the effects.

The paper is organized as follows.
In the next section, we present the governing equations including the EoS.
In section 3, we present the sonic point properties.
In section 4, we present the accretion and wind solutions.
In section 5, we discuss the validity of our relativistic EoS.
Discussion and concluding remarks are presented in the last section.

\section{Assumptions and Equations}

To ensure that the effects of fluid composition are clearly presented,
we keep our model of accretion and wind as simple as possible.
We consider adiabatic, spherical flows onto Schwarzschild black holes.
The space time is described by the Schwarzschild metric
\begin{equation}
ds^2=-\left(1-\frac{2GM_B}{c^2r} \right)c^2dt^2+
\left(1-\frac{2GM_B}{c^2r}\right)^{-1}dr^2
+r^2d{\theta}^2+r^2\sin^2{\theta}d\phi^2,
\end{equation}
where $r$, $\theta$, $\phi$ are the usual spherical coordinates,
$t$ is the time, and $M_B$ is the mass of the central black hole.
Although AGNs and  micro-quasars are in general powered by rotating flows,
studies of spherical flows are not entirely of pedagogic interest.
For instance, such studies can throw light on the nature of accretions onto
isolated black holes in low angular momentum and cold clouds.
In addition, hot spherical flows may mimic accretions very close to
black holes, where the accreting matter is expected to be of low angular
momentum, hot, and with strong advection.
Non-conservative processes and magnetic fields are ignored, too.

The energy-momentum tensor of a relativistic fluid is given by
\begin{equation}
T^{\mu \nu}=(e+p)u^{\mu}u^{\nu}+pg^{\mu \nu},
\end{equation}
where $e$ and $p$ are the energy density and gas pressure
respectively, all measured in the local frame. The four-velocities
are represented by $u^{\mu}$.
The equations governing fluid dynamics are given by
\begin{equation}
T^{\mu \nu}_{; \nu}=0 ~~~~ \mbox{and} ~~~~ (nu^{\nu})_{; \nu}=0,
\end{equation}
where $n$ is the particle number density of the fluid measured in
the local frame.

\subsection{EoS for single-component fluids}

Equation (3) is essentially five independent equations, while the number
of variables are six.
This anomaly in fluid dynamics is resolved by a closure relation between
$e$, $p$ and $n$ (or the mass density $\rho=nm$), and this relation is
known as the EoS.
The EoS for single-component relativistic fluids, which are
{\it in thermal equilibrium}, has been known for a while, and is given by
\begin{mathletters}
\begin{equation}
\frac{e+p}{\rho c^2} = \frac{K_3(\rho c^2/p)}{K_2(\rho c^2/p)}
\end{equation}
\citep{c38,s57}.
Here, $K_2$ and $K_3$ are the modified Bessel functions of the second
kind of order two and three, respectively.

Owing to simplicity, however, the most commonly used EoS has been
the one with a fixed ${\Gamma}$, which is written as
\begin{equation}
e=\rho c^2+\frac{p}{\Gamma -1}.
\end{equation}
As noted in Introduction, this EoS, which admits the superluminal sound
speed, is not applicable to all ranges of temperature \citep{metal05,rcc06}.
Here, we adopt an approximate EoS
\begin{equation}
e = \rho c^2 +p\left(\frac{9p+3\rho c^2}{3p+2\rho c^2}\right),
\end{equation}
which reproduces very closely the relativistically correct EoS
in equation (4a), better than the one proposed by \citet{m71}
\begin{equation}
p = \frac{\rho c^2}{3}\left(\frac{e}{\rho c^2}-\frac{\rho c^2}{e}\right).
\end{equation}
\end{mathletters}
A comparative study of various EoS's for single-component relativistic
fluids was presented in \citet{rcc06}.

\subsection{EoS for multi-component fluids}

We consider fluids which are composed of electrons, positrons, and protons.
Then the number density is given by
\begin{mathletters}
\begin{equation}
n={\Sigma} n_i=n_{e^-}+n_{e^+}+n_{p^+},
\end{equation}
where $n_{e^-}$, $n_{e^+}$, and $n_{p^+}$ are the electron, positron,
and proton number densities, respectively.
Charge neutrality demands that
\begin{equation}
n_{e^-}=n_{e^+}+n_{p^+} ~~~~ \Rightarrow ~~~~ n=2n_{e^-} ~~~~
\mbox{and} ~~~~ n_{e^+}=n_{e^-}(1-\xi),
\end{equation}
where $\xi=n_{p^+}/n_{e^-}$ is the relative proportion of protons.
The mass density is given by
\begin{equation}
\rho=\Sigma n_im_i=n_{e^-}m_e\left\{ 2-\xi
\left(1-\frac{1}{\eta}\right)\right\},
\end{equation}
where $\eta=m_e/m_p$, and $m_e$ and $m_p$ are the electron and proton masses,
respectively.
For {\it single-temperature} fluids, the isotropic pressure is given by
\begin{equation}
p=\Sigma p_i=2n_{e^-}kT.
\end{equation}

As our EoS for multi-component fluids, we adopt
\begin{equation}
e=\Sigma e_i=\Sigma \left[n_im_i c^2 +p_i\left(\frac{9p_i+3n_im_i c^2}
{3p_i+2n_im_i c^2} \right)\right].
\end{equation}
The non-dimensional temperature is defined with respect to the electron
rest mass energy, $\Theta=kT/(m_ec^2)$.
With equations (5a) -- (5d), the expression of the energy density in
equation (5e) simplifies to
\begin{equation}
e=n_{e^-}m_ec^2f,
\end{equation}
where
\begin{equation}
f=(2-\xi)\left[1+\Theta\left(\frac{9\Theta+3}{3\Theta+2}\right)\right]
+\xi\left[\frac{1}{\eta}+\Theta\left(\frac{9\Theta+3/\eta}{3\Theta+2/\eta}
\right)\right].
\end{equation}
The expression of the polytropic index for single-temperature fluids is
given by
\begin{equation}
N=\frac{T}{p}\Sigma n_i\frac{d\Phi_i}{dT}=\frac{1}{2}\frac{df}{d\Theta},
\end{equation}
where
\begin{equation}
\Phi_i=\frac{e_i}{n_i}=m_ic^2+kT\frac{9kT+3m_ic^2}{3kT+2m_ic^2}
\end{equation}
is the energy density per particle of each component.
The effective adiabatic index is calculated by
\begin{equation}
\Gamma=1+\frac{1}{N}.
\end{equation}
The definition of the sound speed, $a$, is
\begin{equation}
\frac{a^2}{c^2}=\frac{\Gamma p}{e+p}=\frac{2 \Gamma \Theta}{f+2\Theta}.
\end{equation}
\end{mathletters}

The polytropic index $N$ (and also the adiabatic index $\Gamma$) is
an indicator of the thermal state of a fluid.
If $N\ \rightarrow\ 3/2$ (or $\Gamma\ \rightarrow\ 5/3$), the fluid is
called thermally non-relativistic.
On the other hand, if $N\ \rightarrow\ 3$ (or $\Gamma\ \rightarrow\ 4/3$),
it is called thermally relativistic.
For single-component fluids, $N$ and $\Gamma$ are given as a function of
the temperature alone \citep{rcc06}.
For multi-component fluids, however, not just the temperature, the mass
of the constituent particles also determines the thermal state.
Hence, the proton proportion, $\xi$, enters as a parameter too.

In Figure 1, we show various thermodynamic quantities and their
inter-relations for fluids with different $\xi$.
In Figure 1a which plots $N$ as a function of $T$, the left most
(solid) curve represents the electron-positron pair fluid ($\xi = 0$)
(hereafter, the $e^- - e^+$ fluid) and the right most (dotted) curve
represents the electron-proton fluid ($\xi = 1$) (hereafter,
the $e^- - p^+$ fluid).
In the $e^- - e^+$ fluid, $N\ \rightarrow\ 3$ for $kT>m_ec^2$, while in
the $e^- - p^+$ fluid, $N\ \rightarrow\ 3$ for $kT>m_pc^2$.
In the intermediate temperature range, $m_ec^2 < kT < m_pc^2$,
$N$ decreases (\ie the fluid becomes less relativistic) with the increase
of $\xi$.
It is because if $\xi$ increases (\ie the proton proportion increases),
the thermal energy required to be in the relativistic regime also increases.
By the same reason, at the same $T$, the local sound speed, $a$, decreases
as $\xi$ increases, as shown in Figure 1b.
However, in Figure 1c, it is shown that the relation between $N$ and $a$
is not as simple as the relation between $N$ and $T$.
At the same $a$, $N$ is smallest for the $e^- - e^+$ fluid, and it
increases and then decreases as $\xi$ increases.
The behavior can be understood as follows.
At the same $a$, as $\xi$ increases, the thermal energy increases,
but at the same time, the rest mass energy increases as well.
As noted in Introduction, it is not the thermal energy, but the
competition between the thermal energy and the rest mass energy that
makes a fluid relativistic.
Consequently, for most values of $a$, $N$ increases for $\xi\ \la\ 0.2$
and then decreases for $\xi\ \ga\ 0.2$.
For very low $a$, $N$ increases up to $\xi\ \sim\ 0.5$, and for very high
$a$, $N$ increases up to $\xi\ \la\ 0.1$.
In summary, {\it at a given temperature, the $e^- - e^+$ fluid is most
relativistic, but at a given sound speed, the $e^- - e^+$ fluid
is least relativistic and fluids with finite proton proportions are
more relativistic}.

\subsection{Equations of motion}

The energy-momentum conservation equation [the first of equation (3)]
can be reduced to the relativistic Euler equation and the entropy equation.
Under the steady state and radial flow assumptions, the equations of motion
are given by
\begin{mathletters}
\begin{equation}
u^r\frac{du^r}{dr}+\frac{1}{r^2}=-\left(1-\frac{2}{r}+u^ru^r\right)
\frac{1}{e+p}\frac{dp}{dr},
\end{equation}
and
\begin{equation}
\frac{de}{dr}-\frac{e+p}{n}\frac{dn}{dr}=0,
\end{equation}
along with the continuity equation [the second of equation (3)]
\begin{equation}
\frac{1}{n}\frac{dn}{dr}=-\frac{2}{r}-\frac{1}{u^r}\frac{du^r}{dr}.
\end{equation}
\end{mathletters}
Here, we use the system of units where $G=M_B=c=1$, so that the units of
length and time are $r_g=GM_B/c^2$ and $t_g=GM_B/c^3$.
It is to be noted that in this system of units, the Schwarzschild radius
or the radius of the event horizon is $r_s=2$.
After some lengthy calculations, equations (6a) -- (6c) are then simplified to
\begin{mathletters}
\begin{equation}
\frac{dv}{dr}=\frac{(1-v^2)[a^2(2r-3)-1]}{r(r-2)(v-a^2/v)}
\end{equation}
and
\begin{equation}
\frac{d\Theta}{dr}=-\frac{\Theta}{N}\left[\frac{2r-3}{r(r-2)}
+\frac{1}{v(1-v^2)}\frac{dv}{dr}\right],
\end{equation}
\end{mathletters}
where the radial three-velocity is defined as $v^2=-u_ru^r/(u_tu^t)$.

For flows continuous along streamlines, equations (7a) -- (7b) admit
the so-called regularity condition, or the critical point condition,
or the sonic point condition \citep{c90} that is given by
\begin{mathletters}
\begin{equation}
a_c=v_c
\end{equation}
and
\begin{equation}
a^2_c=\frac{1}{2r_c-3}.
\end{equation}
Here, $r_c$ is the sonic point location.
Hereafter, the quantities with subscript $c$ denote those at $r_c$.
From equation (5k), we know $a_{\rm max}=1/{\sqrt{3}}$ (also see Figure 1b).
Therefore, from equation (8b), we have $r_c\ {\geq}\ 3$ \citep{bm76}.
Since $dv/dr={\cal N}/{\cal D}\ \rightarrow\ 0/0$ at $r_c$,
$(dv/dr)_{r_c}$ is obtained by the l'Hospital rule
\begin{equation}
\left(\frac{dv}{dr}\right)_{r_c}=\frac{(d{\cal N}/dr)_{r_c}}
{(d{\cal D}/dr)_{r_c}},
\end{equation}
where ${\cal N}$ and ${\cal D}$ are the numerator and denominator of
equation (7a).
The above equation simplifies to
\begin{equation}
{\cal A}\left(\frac{dv}{dr}\right)^2_{r_c}+{\cal B}
\left(\frac{dv}{dr}\right)_{r_c}
+{\cal C}=0,
\end{equation}
where
\begin{equation}
{\cal A}=\left(2+\frac{1-N_ca^2_c+(\Theta_c/\Gamma_c)(d{\Gamma}/d{\Theta})_c}
{N_c(1-a^2_c)} \right)r_c(r_c-2),
\end{equation}
\begin{equation}
{\cal B}=2\frac{1-N_ca^2_c+(\Theta_c/\Gamma_c)(d{\Gamma}/d{\Theta})_c}{N_ca_c},
\end{equation}
and
\begin{equation}
{\cal C}=2\frac{1-N_ca^2_c+(\Theta_c/\Gamma_c)(d{\Gamma}/d{\Theta})_c}
{N_cr_c}-2a^2_c(1-a^2_c).
\end{equation}
Equation (8d) has two roots.
For radial flows, the roots are of the saddle type, where ($dv/dr)_c$
is real and $(dM/dr)_c$ is of opposite signs for the two roots.
Here, $M=v/a$ is the Mach number.
Moreover, the two roots can be either of the acceleration type (A-type),
where $(dv/dr)_c$ is of opposite signs, or of the deceleration type
(D-type), where $(dv/dr)_c$ is negative for both roots.
In the A-type, both the acceleration and wind flows accelerate at
the sonic point.
On the other hand, in the D-type, only the accretion flows
accelerate, while the wind flows decelerate at the sonic point.

By substituting the quantities at the sonic point, equations (7b) give
the temperature gradient at the sonic point
\begin{equation}
\left(\frac{d\Theta}{dr}\right)_{r_c}=-\frac{\Theta_c}{N_c}
\left[\frac{2r_c-3}{r_c(r_c-2)}+\frac{1}{v_c(1-v^2_c)}
\left(\frac{dv}{dr}\right)_{r_c}\right].
\end{equation}
\end{mathletters}
Finally by integrating the equations of motion, we get the relativistic
Bernoulli equation \citep{lpt75}
\begin{equation}
{\cal E}=\frac{(f+2{\Theta})u_t}{(2-\xi+\xi/\eta)},
\end{equation}
where ${\cal E}$ is the Bernoulli parameter or is also known as the
specific energy of flows.
Since we assume adiabatic flows without heating and cooling, ${\cal E}$
is a constant of motion.

\subsection{Procedure to get global solutions}

Combining equation (8b) and (5k) gives $\Theta_c$ in terms of $r_c$
and $\xi$.
Combining it with equation (9) gives a formula involving $r_c$, ${\cal E}$,
and $\xi$ \citep{c90,c96b,fk07}.
If ${\cal E}$ and $\xi$ are given, then $r_c$ is computed from the formula.
Once $r_c$ is known, all the quantities at $r_c$, \eg $\Theta_c$, $v_c$,
$(dv/dr)_{r_c}$,  $(d\Theta/dr)_{r_c}$, etc, are computed from
equations (8a) -- (8h).
Then equations (7a) and (7b) are integrated, starting from $r_c$, once
inwards and then outwards, to obtain the global, transonic solutions
of spherical flows around black holes.
By this way, we can obtain two parameter (${\cal E}$, $\xi$) family
of accretion and wind solutions.

\section{Sonic Point Properties}

In the transonic flows we study, the sonic point plays an important role.
So before we present global solutions in the next section, we first
investigate the properties of the sonic point in this section.
Understanding the sonic-point properties will allow us to have an idea
of the nature of global flow structures.

The sonic point location, $r_c$, that is computed as a function of
${\cal E}$ and $\xi$, is presented in Figure 2a.
Corresponding to each set of ${\cal E}$ and $\xi$ values, there exists a
unique $r_c$.
Each curve, which is given as a function of ${\cal E}$, is for a different
value of $\xi$.
If a flow is more energetic with larger ${\cal E}$, it is characterized
by a smaller value of $r_c$.
However, at the same ${\cal E}$, $r_c$ is smallest for the $e^- - e^+$
fluid (solid line).
The value of $r_c$ increases for $\xi\ \la\ 0.2$, and then starts to
decrease for larger $\xi$.
In other words, if fluids of the same ${\cal E}$ but different $\xi$
are launched at a large distance away from a black hole, then the
$e^- - e^+$ fluid crosses the sonic point closest to the event horizon,
compared to the fluids of finite proton proportion.
Alternatively, at the same $r_c$, ${\cal E}$ is smallest for the
$e^- - e^+$ fluid, and it increases up to $\xi\ \sim 0.2$ and then
decreases for $0.2\ \la\ \xi\ \leq\ 1$.
Although for the same $r_c$ the $e^- - p^+$ (dotted line) is not most
energetic, it is definitely more energetic than the $e^- - e^+$ fluid.
Since fluids of different composition are energetically quite
different at the same $r_c$, or conversely fluids of different
composition but the same ${\cal E}$ form the sonic point at widely
different $r_c$, it is expected that the global solutions of
accretion and wind flows would be quantitatively and qualitatively
different, depending upon the composition of fluids.

In Figures 2b and 2c, we show $T_c$ and $N_c$ as a function of $r_c$.
Equations (8b) tells that the sound speed at the sonic point, $a_c$, is
fixed, once $r_c$ is determined ($a_c$ implicitly depends on ${\cal E}$
and $\xi$ through $r_c$).
So plotting any variable as a function of $r_c$ is equivalent
to plotting it as a function of $a_c$.
As noted above, at the same $a_c$, fluids composed of lighter particles
are colder.
Therefore, in Figure 2b, at the same $r_c$, the temperature is lowest
for the $e^- - e^+$ fluid, and progressively gets higher for fluids
with larger proton proportions, and the maximum temperature is for
the $e^- - p^+$ fluid.
However, as noted before, higher $T_c$ does not necessarily ensure
higher $N_c$ (\ie more relativistic fluids).
In Figure 2c, at the same $r_c$, the $e^- - e^+$ fluid has the lowest $N_c$,
that is, it is least relativistic.
In the range of a few $\la\ r_c\ \la\ 100$, at the same $r_c$, $N_c$
increases as the proton proportion increases for $\xi\ \la 0.2$,
and then starts to decrease for $0.2 \la \xi\ \leq\ 1$.
This is a consequence of the competition between the thermal energy and
the rest mass energy, as discussed in connection with Figures 1c.
In order to make the point even clearer, in Figure 2d, we show $N_c$ as
a function of $\xi$ for a wide range of values of $r_c$.
Each curve with a single value of $r_c$ signifies fluids of different
composition but the same sound speed at the same sonic point.
$N_c$ tends to peak at some values of $\xi$, where the thermal contribution
with respect to the rest mass energy contribution peaks.
For small values of $r_c$ (\ie large $a_c$'s), a small increase of $\xi$
causes the thermal contribution to peak.
For large values of $r_c$ (\ie small $a_c$'s), large proton proportions
are needed to achieve the same.

As discussed in section 2.3, the roots of equation (8d) are either of
the A-type or of the D-type.
At small values of $r_c$, the nature of the sonic point is of the A-type.
It is because if the sonic point form closer to the central object,
the flow is hotter at the sonic point (Figure 2b), and in the wind that
is thermally driven, the flow tends to accelerate at the sonic point.
But beyond a limiting value, say $r_{c{\ell}}$, the nature changes
from the A-type to the D-type, where the wind flow decelerates at the
sonic point.
In Figure 3a, $r_{c{\ell}}$ is plotted as a function of $\xi$.
Since at a given $r_c$ the $e^- - e^+$ fluid is thermally least relativistic,
$r_{c{\ell}}$ is smallest for the fluid.
The limit $r_{c{\ell}}$ increases with $\xi$.
However, since increasing $\xi$ makes fluids `heavy' too,
$r_{c{\ell}}$ peaks around $\xi\ \sim\ 0.75$.
In Figure 3b, we plot the limiting values of ${\cal E}$ corresponding to
$r_{c{\ell}}$, ${\cal E}_{\ell}$, as a function of $\xi$,
such that for ${\cal E} > {\cal E}_{\ell}$ the nature of the sonic point
is of the A-type, for ${\cal E}<{\cal E}_{\ell}$ it is of the D-type.

\section{Spherical Accretion and Wind Solutions}

In this section, we present the global solutions of equations (7a -- 7b)
that were obtained with the procedure described in section 2.4.
In Figure 4, we first compare typical accretion and wind solutions
of the A and D-types for the $e^- - e^+$ fluid.
The solutions of the A-type in the left panels have the sonic point at
$r_c=4$, inside $r_{c{\ell}}$, while those of the D-type in the right
panels have $r_c=30$, beyond $r_{c{\ell}}$ (Figure 3a).
The accretion solutions (solid curves) are characterized by supersonic
flows at the inner boundary and subsonic flows at the outer boundary
(Figures 4c and 4d).
The wind solutions (dotted curves), on the other hand, have subsonic
flows at the inner boundary and supersonic flows at the outer boundary.
The accretion flows around black holes necessarily accelerate inwards.
However, the wind flows may accelerate (Figure 4a) or decelerate
(Figure 4b) outwards.
The wind solutions considered in this paper are thermally driven.
These winds are very hot at the base, and are powered by the conversion
of the thermal energy into the kinetic energy.
It can be shown from equation (7b) that
\begin{equation}
-\frac{d{\Theta}}{dr}{\leq}\frac{\Theta}{N}\left[ \frac{2r-3}{r(r-2)}\right]
\hskip 0.5cm
{\Rightarrow} \hskip 0.5cm \frac{dv}{dr}{\leq}0.
\end{equation} 
In other words, if the outward thermal gradient is weaker than the gravity,
the outflow can decelerate.
For the wind with $r_c=30$ (Figures 4b, 4d, and 4f), $-{d\Theta}/dr\ \sim$
$(\Theta/N) (2r-3) / [r(r-2)]$ at $r \sim 9.16$, exactly where the outflow
starts to decelerate.
However, the wind velocity will reach an asymptotic value at $r\ \rightarrow$
large, since $-{d\Theta}/dr\ \sim$ $(\Theta/N) (2r-3) / [r(r-2)]\ \sim\ 0$
at large distances from the black hole.
Similar relation between the gradients at the sonic point will determine
the nature of the sonic point.
It may be noted that at $r_c\ \geq\ r_{c{\ell}}$ (Figures 3a and 3b), such
relation between $\left(dv/dr\right)_c$ and $\left(d{\Theta}/dr\right)_c$
is satisfied.
Regardless of accretion/wind or the type, the temperature decreases
with increasing $r$ (Figures 4e and 4f).

We note that the winds in our solutions are too weak to be the precursor
of astrophysical jets, until and unless other accelerating processes like
those caused by magnetic fields or disc radiation are considered \citep{c05}.
In fact, we checked that it is not possible to generate the terminal
speed much greater than $\sim\ 0.8c$ for purely thermally driven winds,
such as the ones that are considered in this paper.
It is also to be noted that our D-type, wind solution is not an example
of `breeze'.
A breeze is always subsonic, while the wind here is transonic, albeit
decelerating.

In the previous figure, we have compared the solutions with the same
$\xi$ (= 0) but different $r_c$.
In Figure 5, we compare the solutions with the same $r_c$ (= 20) but
different $\xi$.
As shown in Figure 2a, even for the same $r_c$, the specific energy is
different for fluids of different $\xi$.
Furthermore, the polytropic index at the sonic point is different too
($N_c=1.547$ in Figure 5a, $N_c=2.626$ in Figure 5b, and $N_c=2.271$ in
Figure 5c).
Therefore, even if we fix the sonic point (and therefore $a_c$), the flow
structure and energetics are different for fluids with different $\xi$.
In these particular solutions, the $e^- - e^+$ fluid is not hot enough to
drive an accelerating wind (Figure 5a), while the fluids with
significant protons can do so.
As in the previous case of decelerating wind solution (\ie Figure 4b),
in the present case the $e^- - e^+$ fluid first accelerates and then starts
to decelerate at $r\ \sim\ 9.86$.
The velocity profile eventually tapers off to an asymptotic value at
large distances away from the black hole.
It has been shown in Figure 2b that at the same $r_c$, adding protons
increases the temperature at the sonic point.
Larger temperature gradient causes winds of finite proton proportion
to be accelerated at the sonic point (Figure 5b).
It is seen that beyond a critical value, the increase in $\xi$ increases
the inertia which reduces the wind speed, as is vindicated by Figures 5b
and 5c.
It is to be remembered that, the D type sonic point is a reality for
fluids of any $\xi$, provided $r_c \ga r_{c \ell}$.

Although the wind solutions are noticeably different depending on $\xi$,
there seems to be only small difference in the velocity profile of the
accretion solutions.
Henceforth, we concentrate only on accretion solutions.
Such small difference in $v$ in accretion solutions is expected.
The accretion is generated mostly by the inward pull of the gravity,
which gives the unique inner boundary condition for black holes,
\ie $v=1$ at $r=2$, regardless of other considerations.
The pressure gradient changes the profile of $v$ too.
Since the composition of fluids determines the thermal state, it
influences the profile of $v$, but the effect is not the dominant one.

In Figure 6, we compare the accretion solutions with the same ${\cal E}$
(= 1.015) but different $\xi$.
As noted below equation (9), ${\cal E}$ is a constant of motion.
For $r\ \rightarrow\ \infty$, as $u_t\ \rightarrow\ 1$, we have
${\cal E}\ \rightarrow\ h_\infty$, where
\begin{equation}
h_{\infty} = \left[\frac{e+p}{\rho}\right]_{\infty}
= \left[\frac{f+2\Theta}{2-\xi+\xi/\eta}\right]_{\infty}
\end{equation} 
is the specific enthalpy at infinity.
Equation (11) tells us that at large distances from black holes,
for the same ${\cal E}$, $T$ is large if $\xi$ is large.
Hence, fluids with larger $\xi$ are hotter to start with.
Therefore even for fluids with the same ${\cal E}$, the solutions are
different if $\xi$ is different.
Figure 6a shows the velocity profile as a function of $r$. 
Here, the difference in $v$ for fluids with different $\xi$ is evident,
albeit not big as pointed above.
Figures 6b, 6c and 6d show the mass density, temperature, and polytropic
index.
To compute the mass density, we need to supply the mass accretion rate,
which is given as
\begin{equation}
{\dot M}=4\pi r^2 u^r \rho
\end{equation}
from equation (6c).
The mass density in Figure 6b was computed for $M_B = 10M_{\odot}$ and
${\dot M}=0.1{\dot M}_{\rm Edd}$, where ${\dot M}_{\rm Edd}$ is the
Eddington rate of accretion.
The difference in $T$ and $N$ for fluids with different $\xi$ is
more pronounced.
The $e^- - e^+$ fluid is slowest, densest (for the same ${\dot M}$),
coldest, and least relativistic.
The $e^- - p^+$ fluid is more relativistic than the $e^- - e^+$ fluid.
But the most relativistic fluid is the one with the intermediate value
of $\xi$.
It is interesting to note that except for the $e^- - e^+$ fluid,
$N$ is a slowly varying function of $r$ for the other two fluids.
Does this mean it would be sufficient to adopt the fixed ${\Gamma}$
EoS with appropriate values of $\Gamma$?

Finally in Figure 7, we compare the accretion solutions with the same
temperature at large distances but different $\xi$.
All the fluids start with $T = T_{\rm out} = 1.3 \times 10^9$ K at
$r = r_{\rm out} =2000$.
Again the mass density was computed for $M_B = 10M_{\odot}$ and
${\dot M}=0.1{\dot M}_{\rm Edd}$.
It is to be noted that the fluids starting with the same $T_{\rm out}$ but
different $\xi$ have different specific energies.
Hence, the velocity at the outer boundary is different too.
As shown in Figure 7a, in these particular solutions, the $e^- - e^+$
fluid starts with a velocity substantially different from those of the other
two fluids, so the resulting velocity profile is substantially different.
From Figure 7d, it is clear that there are significant variations in $N$
for all the fluids.
The $e^- - e^+$ fluid starts with the largest $N$.
It is because at the same temperature, the $e^- - e^+$ fluid is
thermally most relativistic.
The behavior of $N$ can be traced back to Figure 1a.
For instance, the variations in $N$ tend to flatten at $T\ \ga 10^{10}$ K.
In Figure 7c, for the fluids with $\xi=0.5$ and 1, $T\ \la 10^{10}$ K
for $r\ \ga\ 100$ and $T\ \ga 10^{10}$ K for $r\ \la\ 100$.
So significant variations are expected in $N$ at $r{\ga}100$, while the
variations flatten at $r<100$.
Similar considerations will explain the variations in $N$ for the
$e^- - e^+$ fluid.
From Figure 7d, it is clear that we need to adopt a relativistically
correct EoS [equation (4c) or (4d)], instead of the EoS with a fixed
$\Gamma$, in order to capture the proper thermal properties of flows
around black holes.

In this section, we have shown that fluids with different composition
can result in dramatically different accretion and wind flows, even if
they have the same sonic point or the same specific energy, or they
start with the same temperature at large distances from black holes.
So not just adopting a correct EoS, but incorporating the effects of
fluid composition into the EoS (see equation 5e) should be also important
in describing such flows.

\section{Validity of EoS} 

In section 2, we have made the following assumptions for our EoS
(equation 5e); fluids are in equilibrium, \ie 1) the distribution of
the constituent particles is relativistically Maxwellian and 2) the
multi-components are of single temperature.
However, it is not clear whether the conditions are satisfied.
Most astrophysical fluids, unlike the terrestrial ones, consist of
charged particles, which are {\it collisionless}, and so held together
by magnetic fields.
The constituent particles, on the other hand, exchange energies, and
become relaxed mostly through the Coulomb interaction, which is a slow
process in collisionless plasmas.
In addition, most of the heating processes, such as viscosity and shock
heating, are likely to affect protons.
However, it is mainly the electrons which radiate.
So the energy exchange between electrons and protons should operate,
and eventually govern the thermal properties of fluids.

Let $t_{\rm ee}$ be the electron-electron relaxation time scale,
$t_{\rm pp}$ be the proton-proton relaxation time scale, and
$t_{\rm ep}$ be the electron-proton relaxation time scale.
And let $t_{\rm prob}$ be the time scale of problem, such as the dynamical
time scale, or the heating and/or cooling time scale.
Only if $t_{\rm ee} < t_{\rm prob}$ and $t_{\rm pp} < t_{\rm prob}$,
electrons and protons will separately attain the Maxwellian distributions.
And only if $t_{\rm ep} < t_{\rm prob}$, electrons and protons will relax
to single temperature.

To verify the assumptions for our EoS, in this section, we compare the
relaxation time scales with the dynamical or accretion time scale
($t_{\rm dyn} = r/v$) for an accretion solution.
We consider the temperature range where protons are thermally
non-relativistic while electrons are relativistic.
In most our solutions in the previous section, the computed temperature
favors this range.
The relativistic electron-electron interaction time scale was derived
by \citet{s83},

\begin{mathletters}
\begin{equation}
t_{\rm ee}=\frac{8k^2}{(m_ec^2)^2{\sigma}_Tc{\rm ln}{\Lambda}}
\frac{T^2}{n_{e^-}}.
\end{equation}
The time scale for the non-relativistic proton-proton interaction is given
in \citet{s62},
\begin{equation}
t_{\rm pp}=\frac{4{\sqrt{\pi}}k^{3/2}}{{\rm ln}{\Lambda}(m_pc^2)^{3/2}
{\sigma}_Tc} \left(\frac{m_p}{m_e}\right)^2\frac{T^{3/2}}{n_{p^+}}.
\end{equation}
The relativistic electron-proton interaction time scale was also derived
by \citet{s83},
\begin{equation}
t_{\rm ep}=2\left(\frac{m_p}{m_e}\right)\left(\frac{\kappa}{m_ec^2}\right)
\frac{1}{\sigma_Tc}\frac{T}{n_{p^+}}
\end{equation}
\end{mathletters}

We present the electron number density, $n_{e^-}$ (Figure 8a), the three
velocity $v$ (Figure 8b) and the temperature $T$ (Figure 8c) of the
accretion solution for the $e^- - p^+$ fluid with ${\cal E} = 1.5247$.
The electron number density was computed for $M_B = 10M_{\odot}$ and
${\dot M}=0.1{\dot M}_{\rm Edd}$.
In Figure 8d, various time scales are compared.
All the relaxation time scales were calculated for the solution of
single-temperature.
To our surprise, it is clear that the accretion flow in the figure is
`too fast', such that various relaxation time scales are longer than
the accretion time scale at least within few tens of $r_s$.
The implication of it is not clear, however.
For instance, in relativistic plasmas, the constituent particles can be
relaxed through the interactions with magnetic fields, too.
But the relaxation will depend on the details of field configuration,
such as the strength and the topology.
Since we ignore in this study magnetic fields as well as other processes
such as non-conservative ones, we leave this issue of the validity of our
EoS for future studies.

\section{Discussion and Concluding Remarks}

In this paper, we have investigated the effects of fluid composition on
the solutions of accretion and wind flows onto black holes.
In order to elucidate the effects, we have considered a very simple model
of spherical flows onto Schwarzschild black holes, and non-conservative
processes and magnetic fields have been ignored.

First, we have suggested an approximate EoS for multi-component fluids in
equation (5e), and studied the thermal properties of fluids with the EoS.
Three temperature ranges have been categorized;
for $kT < m_ec^2$, any type of fluids are thermally non-relativistic,
for $kT > m_pc^2$, any type of fluids are thermally relativistic, and
for $m_ec^2 < kT < m_pc^2$, the degree to which fluids are relativistic
is determined by the composition of the fluids as well as the temperature
(Figure 1a).
Then we have shown that although at the same temperature the $e^- - e^+$
fluid is most relativistic (Figure 1a), at the same sound speed it is
least relativistic (Figure 1c), compared to the fluids with protons.
It is because whether a fluid is relativistic or not depends on the
competition between the thermal energy and the rest mass energy of
the fluid.

The thermal properties of fluids carry to the sonic point properties.
The sound speed at the sonic point, $a_c$, explicitly depends only
on the sonic point location, $r_c$ (it implicitly depends on the
specific energy, ${\cal E}$, and the proton proportion, $\xi$,
through $r_c$).
Therefore, comparing the thermodynamic quantities at the same $r_c$ is
equivalent to comparing those quantities at the same $a_c$.
We have shown that at the same $r_c$, the $e^- - e^+$ fluid is least
relativistic, and a fluid with a finite $\xi$ is most relativistic
(Figures 2c and 2d).

Then, we have presented the global solutions of accretion and wind flows
for the same $r_c$ but different $\xi$, for the same ${\cal E}$ but
different $\xi$, and for the same $T$ at large distances from black holes
but different $\xi$.
In all the cases, the flows can be dramatically different, if the
composition is different.
This asserts that the effects of fluid composition are important
in the solutions, and hence, incorporating them properly into the
solutions through the EoS is important.

Lastly, we have noted that the EoS in equation (5e) is based on the
assumptions that the distribution of the constituent particles is
relativistically Maxwellian and the multi-components are of single
temperature.
However, at the same time, we have pointed out that while the Coulomb
relaxation times are normally shorter than the dynamical time far away
from black holes, close to black holes they can be longer.
It means that close to black holes, the assumptions for the EoS can
be potentially invalidated.
The implication of it needs to be understood, and we leave further
consideration of this issue for future studies.

\acknowledgments

The work of DR was supported by the Korea Research Foundation Grant
funded by the Korean Government (MOEHRD) (KRF-2007-341-C00020).

\clearpage

\begin{figure}
\hskip -1.3cm
\includegraphics[scale=0.32]{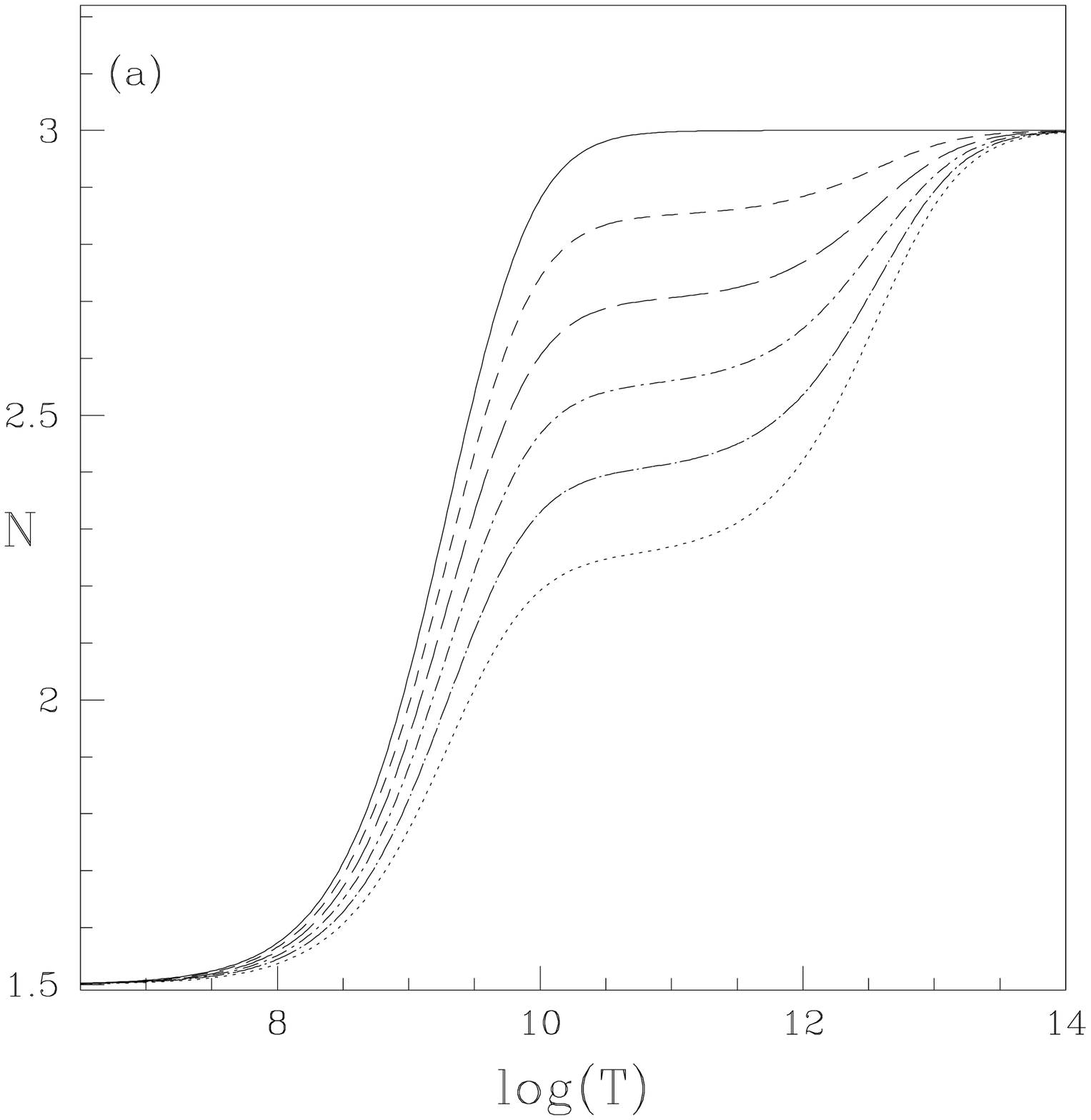}
\vskip -6.5cm
\hskip 4.8cm
\includegraphics[scale=0.32]{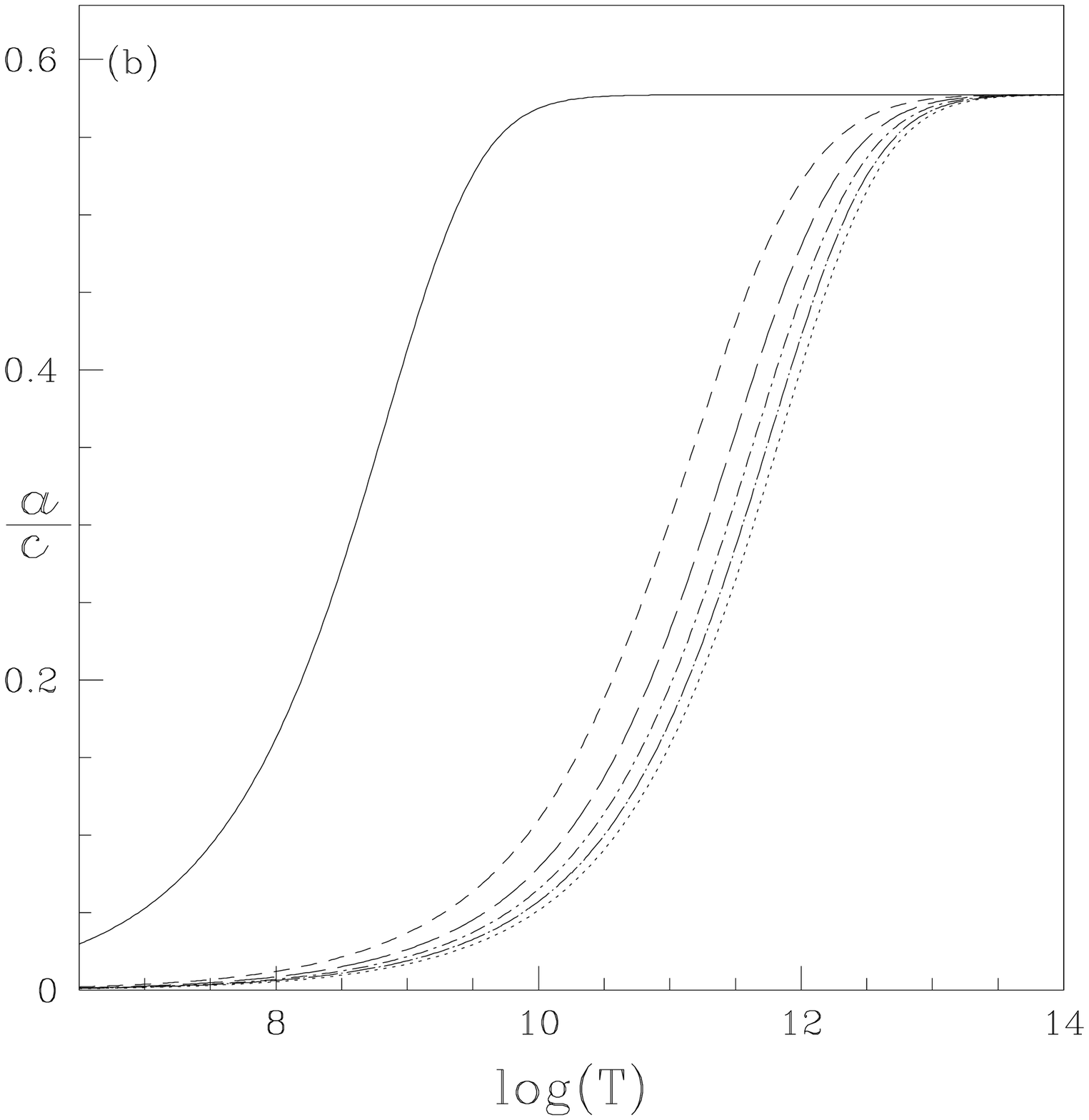}
\vskip -6.5cm
\hskip 10.8cm
\includegraphics[scale=0.32]{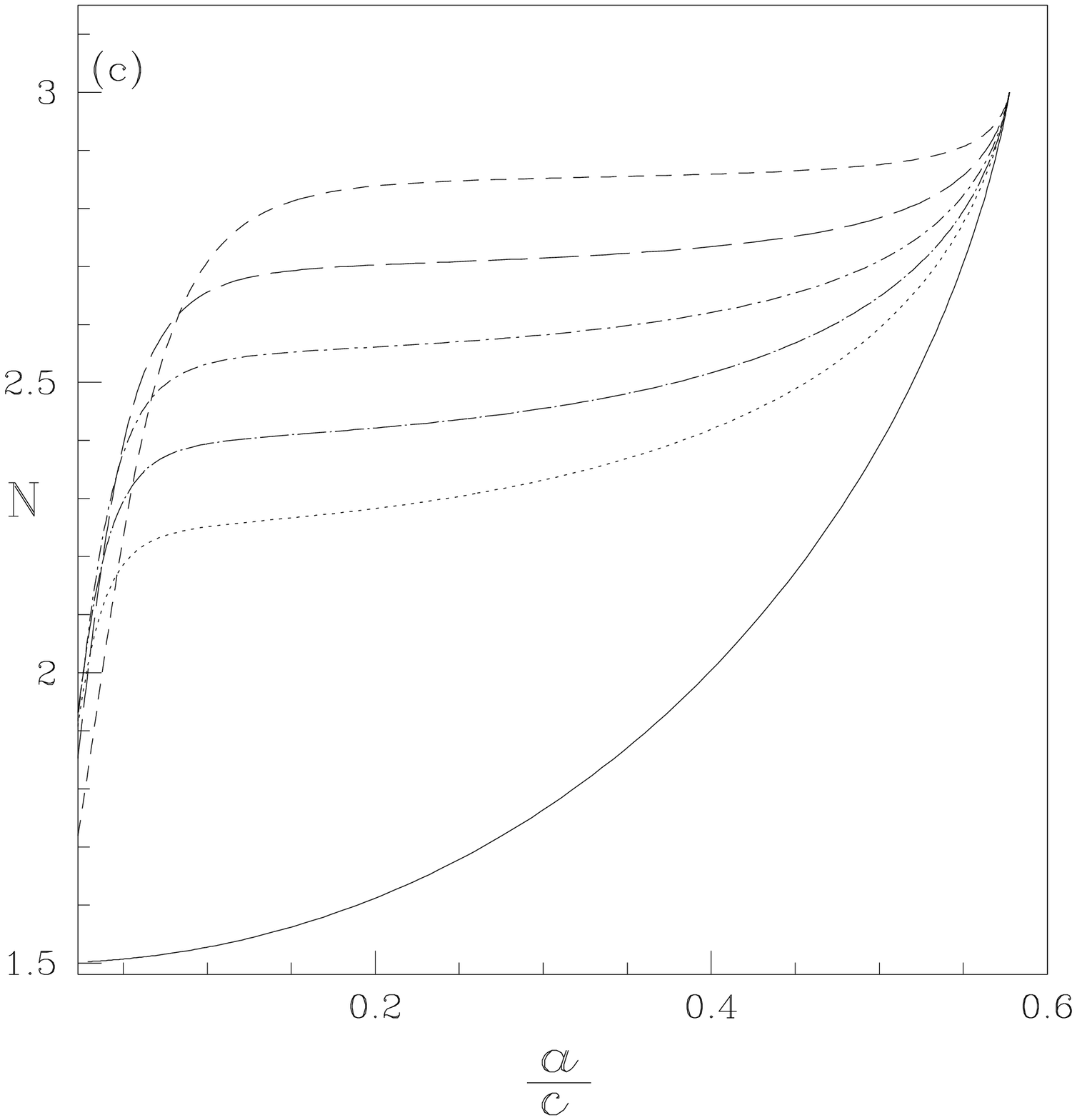}
\vskip 1.0cm
\caption{(a) Polytropic index as a function of the temperature,
(b) sound speed as a function of the temperature, and
(c) polytropic index as a function of the sound speed,
for multi-component relativistic fluids of different composition with
proton proportions of $\xi=0$ (solid), $0.2$ (dashed), $0.4$
(long dashed), $0.6$ (dashed-dotted), $0.8$ (long dashed-dotted),
and $1$ (dotted).
Hereafter, the temperature in figures is given in units of Kelvin.}
\end{figure}

\clearpage

\begin{figure}
\hskip -0.4cm
\includegraphics[scale=0.8]{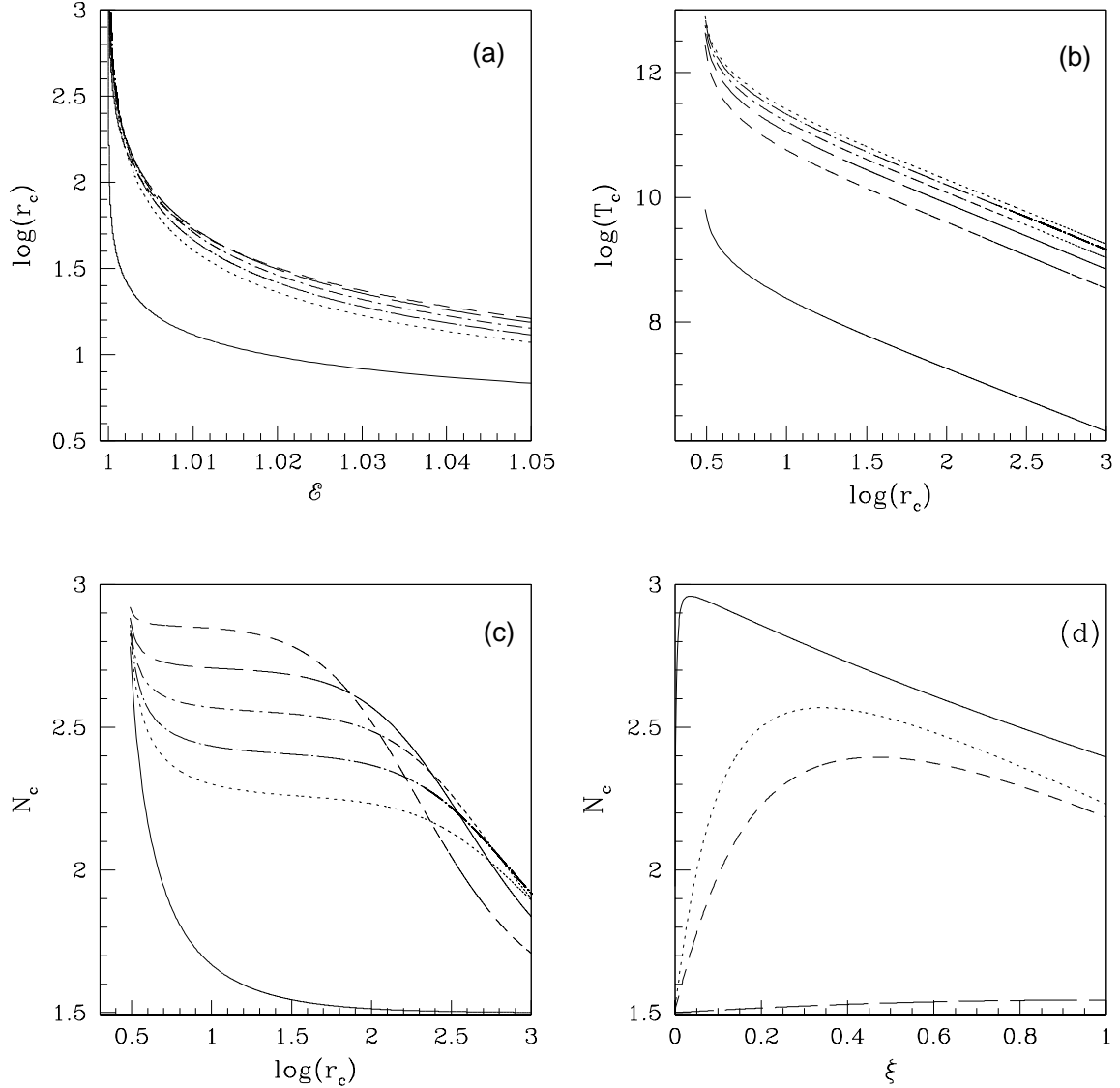}
\vskip -0.7cm
\caption{(a) Sonic point location as a function of the specific energy,
(b) temperature at the sonic point as a function of the sonic point
location,
(c) polytropic index at the sonic point as a function of the sonic point
location,
for transonic flows of fluids with $\xi=0$ (solid), $0.2$ (dashed),
$0.4$ (long dashed), $0.6$ (dashed-dotted), $0.8$ (long dashed-dotted),
and $1$ (dotted).
(d) Polytropic index at the sonic point as a function of proton
proportion for the flows with the sonic point location of $r_c=5$ (solid),
$105$ (dotted), $205$ (dashed) and $13505$ (long dashed).}
\end{figure}

\clearpage

\begin{figure}
\hskip -0.6cm
\includegraphics[scale=0.45]{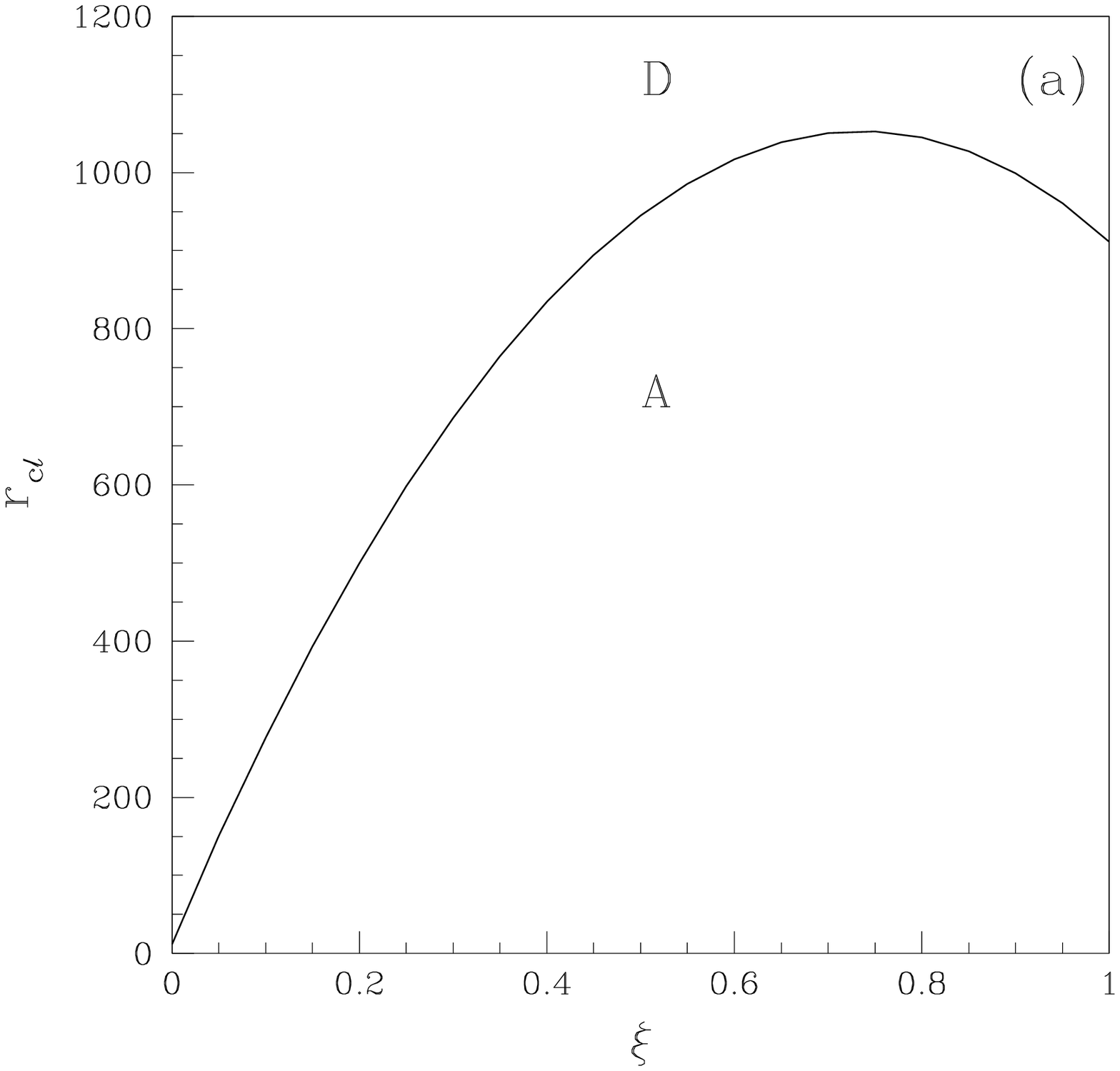}
\hfil
\includegraphics[scale=0.45]{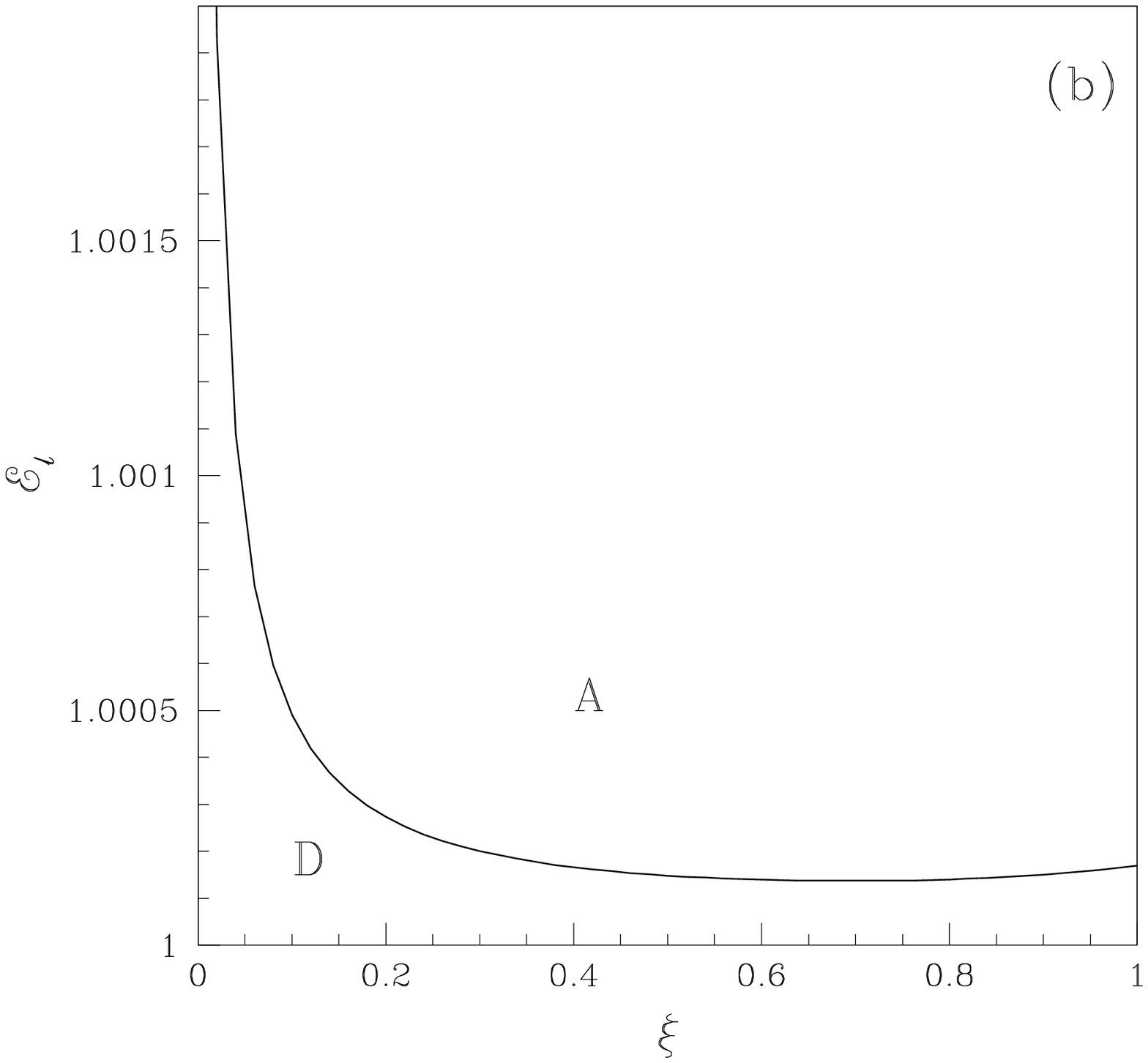}
\vskip -0.2cm
\caption{Limiting values of (a) the sonic point location and (b) the
specific energy, which divide the domain of the A-type sonic point roots
from that of the D-type sonic point roots, as a function of proton
proportion.}
\end{figure}

\clearpage

\begin{figure}

\vskip 0.5cm
\hskip -0.4cm
\includegraphics[scale=0.8]{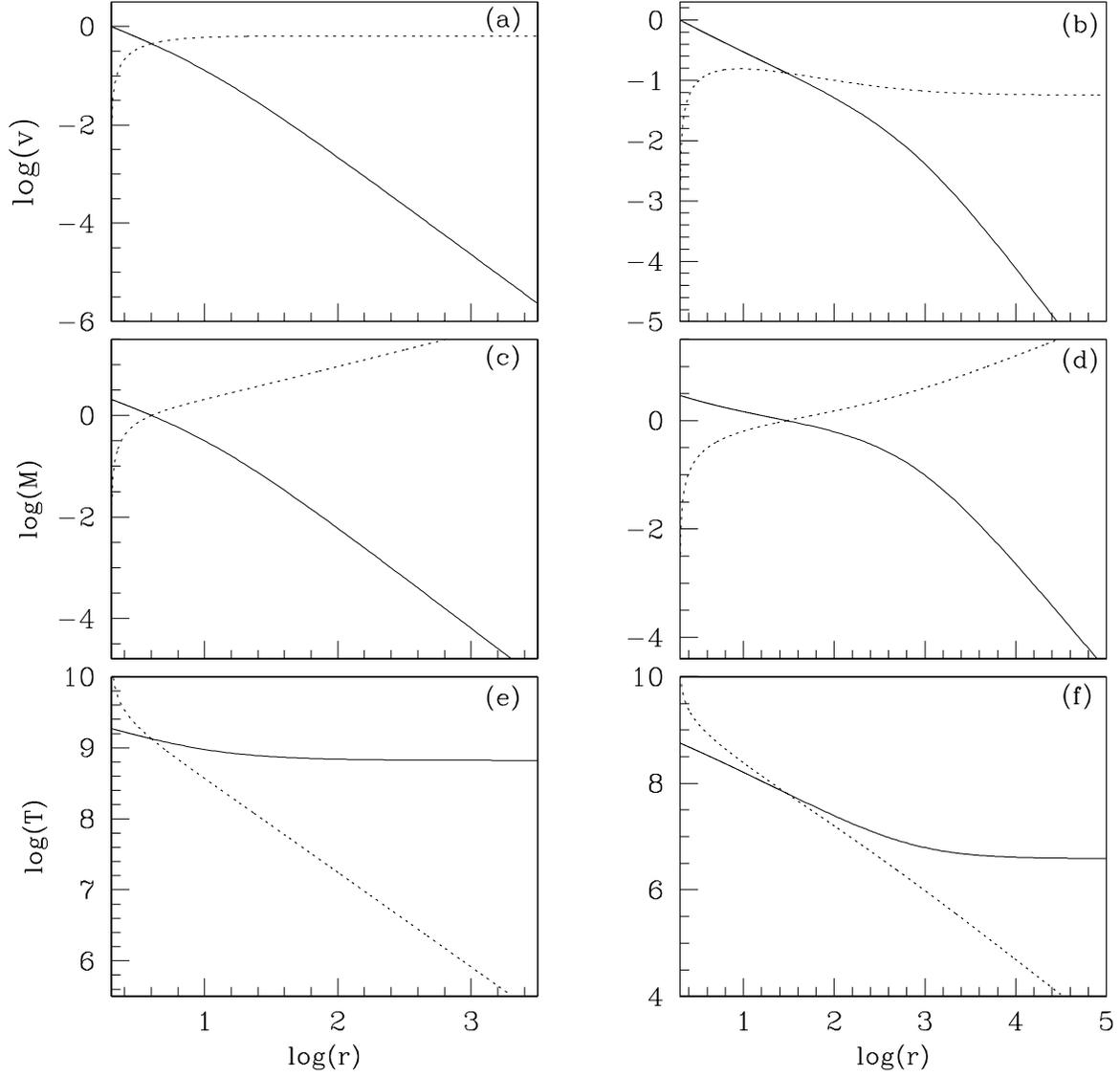}
\vskip -0.7cm
\caption{Examples of transonic accretion (solid) and wind (dotted)
solutions of the A-type with the sonic point at $r_c=4$
(${\cal E}=1.3$, $N_c=2.163$) ({\it left panels}) and of the D-type
with the sonic point at $r_c=30$ (${\cal E}=1.0016$, $N_c=1.548$)
({\it right panels}).
The $e^- - e^+$ fluid ($\xi=0$) is considered.
The radial three-velocity ({\it top}), Mach number ({\it middle}), and
temperature ({\it bottom}) are shown as a function of radius.}
\end{figure}

\clearpage

\begin{figure}
\hskip -2.2cm
\includegraphics[scale=0.4]{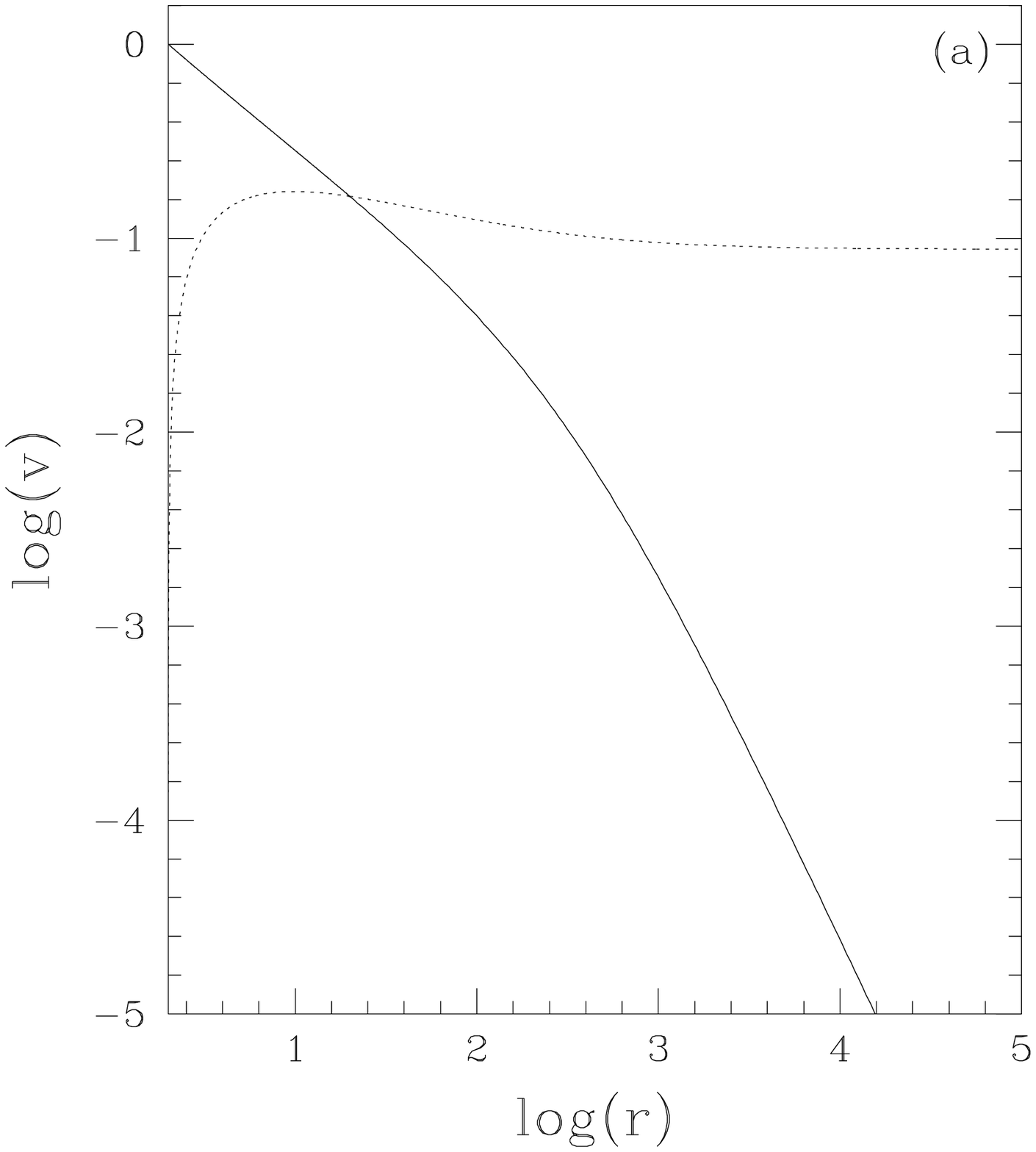}
\vskip -8.13cm
\hskip 3.8cm
\includegraphics[scale=0.4]{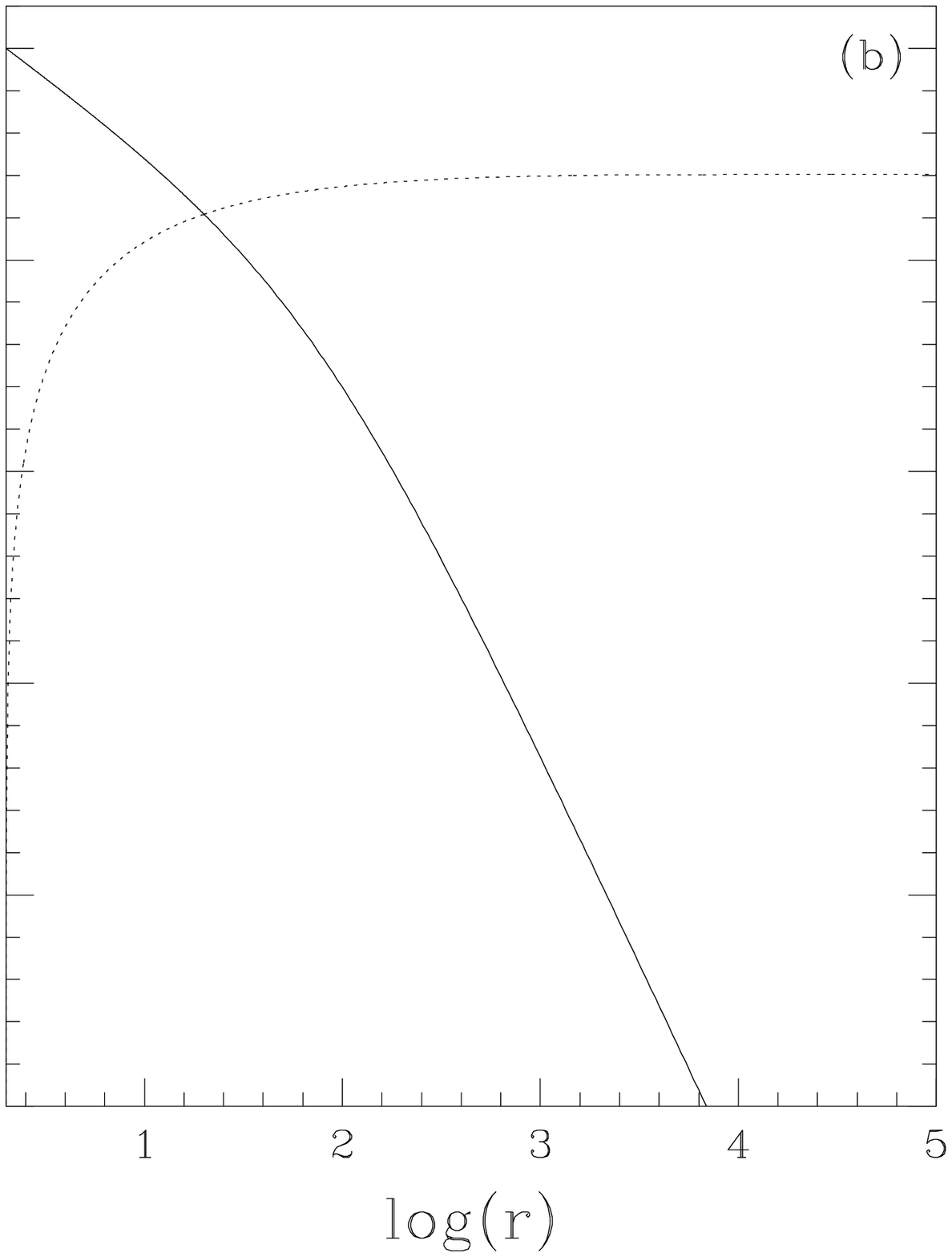}
\vskip -8.13cm
\hskip 9.85cm
\includegraphics[scale=0.4]{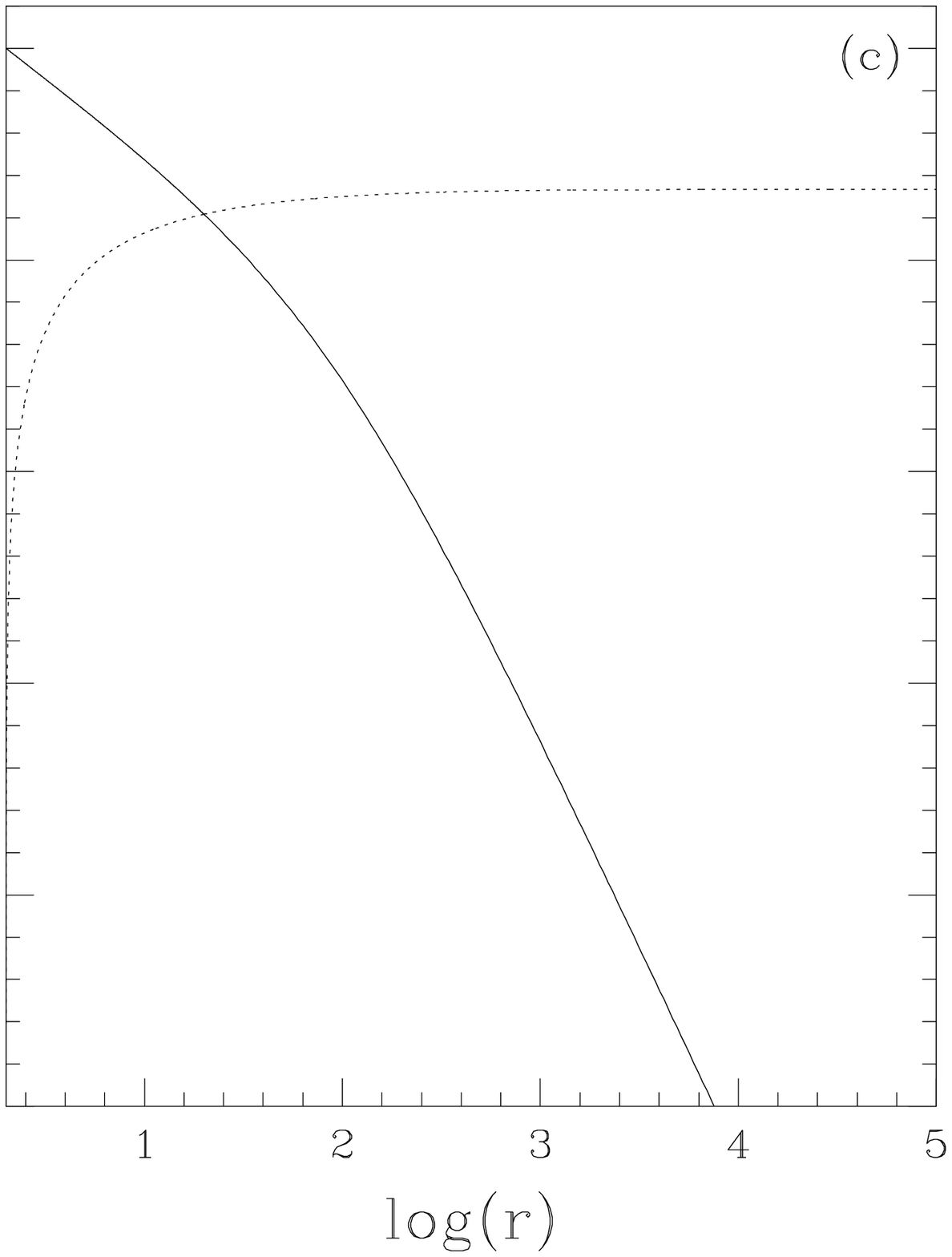}
\caption{Comparison of accretion (solid) and wind (dotted) solutions
with the same sonic point location of $r_c=20$ but different proton
proportions of (a) $\xi=0$ (${\cal E}=1.0039$), (b) $\xi=0.5$
(${\cal E}=1.0337$), and (c) $\xi=1$ (${\cal E}=1.0239$).
The radial three-velocity is shown as a function of radius.}
\end{figure}

\clearpage

\begin{figure}
\includegraphics[scale=0.8]{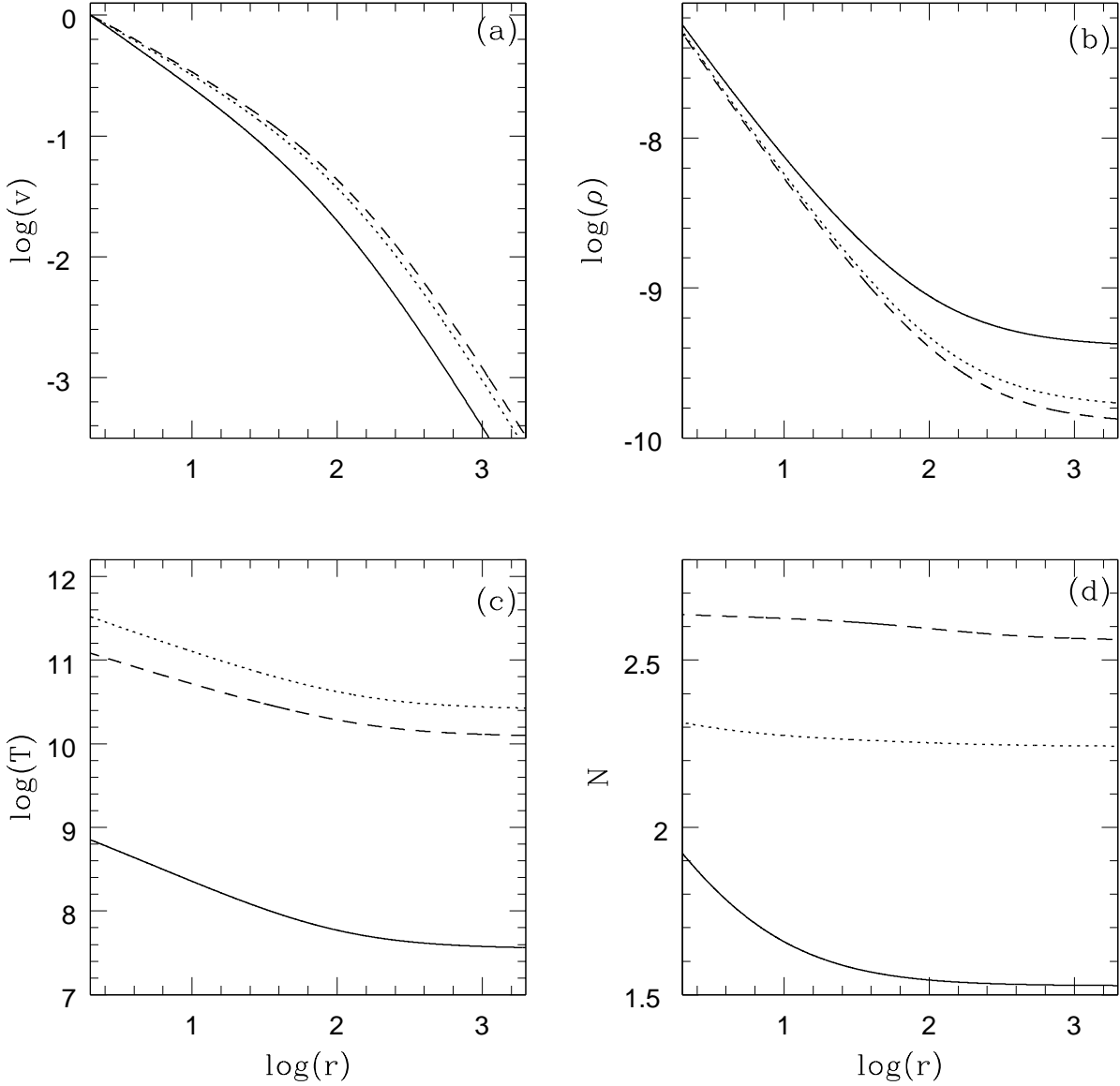}
\vskip -0.5cm
\caption{Comparison of accretion solutions with the same specific energy
of ${\cal E}=1.015$ but different proton proportions of $\xi=0$ (solid),
$\xi=0.5$ (dashed), and $\xi=1$ (dotted).
The sonic point locations are $r_c=11.0$ for $\xi=0$, $r_c=38.267$ for
$\xi=0.5$, and $r_c=28.972$ for $\xi=1$.
The radial three-velocity (a), mass density (b), temperature (c), and
polytropic index (d) are shown as a function of radius.
The mass density was computed assuming the black hole of
$M_B=10M_{\odot}$ and the accretion rate of
${\dot M}=0.1{\dot M}_{\rm Edd}$, and is given in units of g cm$^{-3}$.}
\end{figure}

\clearpage

\begin{figure}
\includegraphics[scale=0.8]{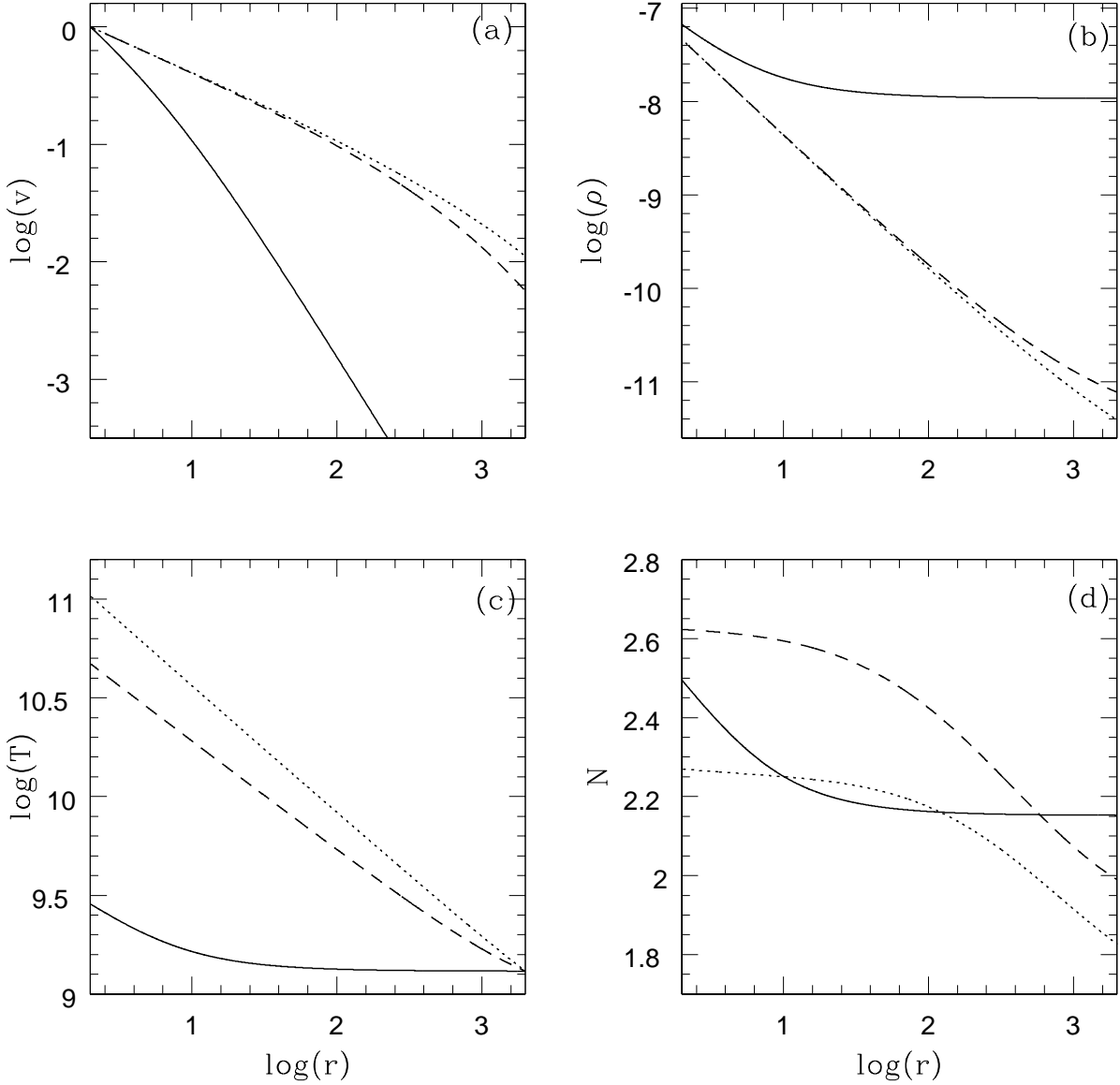}
\vskip -0.5cm
\caption{Comparison of accretion solutions with the same temperature of
${T}_{\rm out}=1.3{\times}10^9$ K at the outer boundary $r_{\rm out}=2000$
but different proton proportions of $\xi=0$ (solid), $\xi=0.5$ (dashed),
and $\xi=1$ (dotted).
The sonic point location and the specific energy are $r_c=3.5$ and
${\cal E}=1.6322$ for $\xi=0$, $r_c=333.3$ and ${\cal E}=1.0008$ for
$\xi=0.5$, and $r_c=806.4$ and ${\cal E}=1.0002$ for $\xi=1$.
The radial three-velocity (a), mass density (b), temperature (c), and
polytropic index (d) are shown as a function of radius.
The mass density was computed assuming the black hole of
$M_B=10M_{\odot}$ and the accretion rate of
${\dot M}=0.1{\dot M}_{\rm Edd}$, and is given in units of g cm$^{-3}$.}
\end{figure}

\clearpage

\clearpage

\begin{figure}
\begin{center}
\includegraphics[scale=0.7]{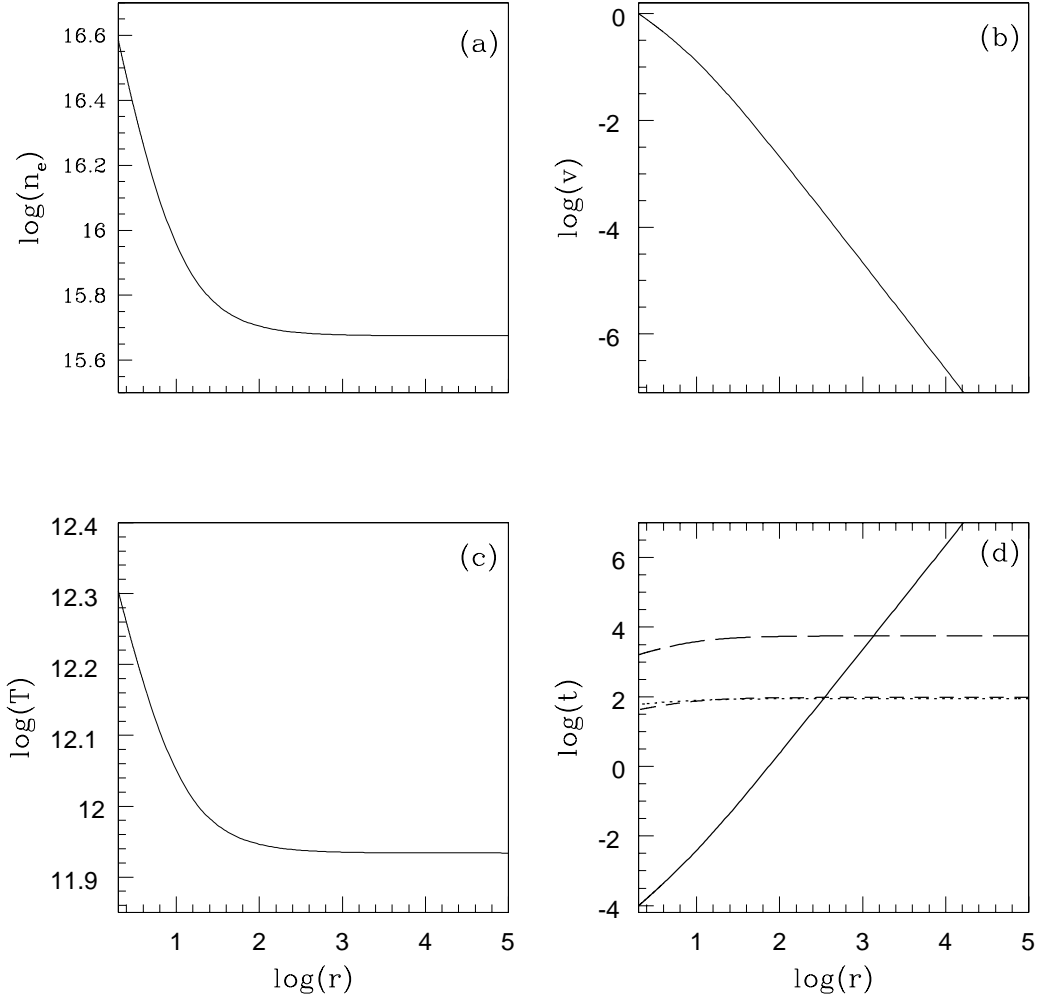}
\end{center}
\caption{(a) Electron number density, (b) radial three-velocity, and (c)
temperature as a function of radius in an accreting flow of the $e^- - p^+$
fluid ($\xi=1$) with ${\cal E}=1.5247$.
${\dot M}=0.1{\dot M}_{\rm Edd}$ onto a black hole of $M_B=10M_{\odot}$
was assumed.
(d) Comparison of various time scales of the same flow.
Different curves represent the accretion time $t_{\rm dyn}$ (solid),
the electron-electron relaxation time $t_{\rm ee}$ (dotted),
the proton-proton relaxation time $t_{\rm pp}$ (dashed),
and the electron-proton relaxation time $t_{\rm ep}$ (long-dashed).}
\end{figure}

\end{document}